%% file: conference_101719.tex
\documentclass[conference,a4paper]{IEEEtran}
\IEEEoverridecommandlockouts
\usepackage{cite}
\usepackage{amsmath,amssymb,amsfonts}
\usepackage{algorithmic}
\usepackage{graphicx}
\usepackage{enumitem}
\usepackage{textcomp}
\usepackage{xcolor}
\usepackage[capitalise,noabbrev]{cleveref}
\usepackage{lazyeqn}
\usepackage{comment}
\usepackage{url}
\usepackage{booktabs}
\usepackage{siunitx}
\input{acronyms}

\makeatletter
\newcommand*{\org@overidelabel}{}
\let\org@overridelabel\@verridelabel
\@ifpackagelater{acronym}{2015/03/21}{
  \renewcommand*{\@verridelabel}[1]{%
    \@bsphack
    \protected@write\@auxout{}{\string\AC@undonewlabel{#1@cref}}%
    \org@overridelabel{#1}%
    \@esphack
  }%
}{
  \renewcommand*{\@verridelabel}[1]{%
    \@bsphack
    \protected@write\@auxout{}{\string\undonewlabel{#1@cref}}%
    \org@overridelabel{#1}%
    \@esphack
  }%
}
\makeatother

\newtheorem{proposition}{Proposition}
\newtheorem{fact}{Fact}
\DeclareMathOperator{\spn}{span}
\DeclareMathOperator*{\minimize}{minimize}

\begin{document}

\title{Digital Post-Distortion Architectures for Nonlinear Power Amplifiers: Volterra and Kernel Methods
\thanks{The authors acknowledge the financial support by the Federal Ministry of Education and Research of Germany (BMBF) in the project 6G-ANNA (project identification number: 16KISK087) and the joint project 6G-RIC in the programme of “Souverän. Digital. Vernetzt.” (project identification numbers: 16KISK020K, 16KISK030). The authors alone are responsible for the content of the paper.}}

\author{\IEEEauthorblockN{Daniel Sch\"aufele,
Jochen Fink, Renato L. G. Cavalcante,
and S{\l}awomir Sta\'nczak}
\IEEEauthorblockA{Wireless Communications and Networks,
Fraunhofer Heinrich Hertz Institute, Berlin, Germany\\
\{daniel.schaeufele, jochen.fink, renato.cavalcante, slawomir.stanczak\}@hhi.fraunhofer.de}
}
\maketitle

\begin{abstract}
In modern 5G \acp{UE}, the \ac{PA} contributes significantly to power consumption during uplink transmissions, especially in cell-edge scenarios.
While reducing power backoff can enhance \ac{PA} efficiency, it introduces nonlinear distortions that degrade signal quality.
Existing solutions, such as digital pre-distortion, require complex feedback mechanisms for optimal performance, leading to increased \ac{UE} complexity and power consumption.
Instead, in this study we explore \ac{DPoD} techniques, which compensate for these distortions at the base station, leveraging its superior computational resources.

In this study, we conduct an comprehensive study concerning the challenges and advantages associated with applying \ac{DPoD} in time-domain, frequency-domain, and DFT-s-domain.
Our findings suggest that implementing \ac{DPoD} in the time-domain, complemented by frequency-domain channel equalization, strikes a good balance between low computational complexity and efficient nonlinearity compensation.
In addition, we demonstrate that memory has to be taken into account regardless of the memory of the \ac{PA}.
Subsequently, we show how to pose the complex-valued problem of nonlinearity compensation in a real Hilbert space, emphasizing the potential performance enhancements as a result.
We then discuss the traditional Volterra series and show an equivalent kernel method that can reduce algorithmic complexity.
Simulations validate the results of our analysis and show that our proposed algorithm can significantly improve performance compared to state-of-the-art algorithms.
\end{abstract}
\acresetall


\section{Introduction}

In commercially available 5G \acp{UE}, the \ac{PA} consumes most of the power of uplink transmissions in cell-edge scenarios\cite{schippers2024data}.
For this reason optimizing the \ac{PA} energy-efficiency is of utmost importance.
Generally, the lower the power backoff, the more efficiently the \ac{PA} will operate.
However, a low power backoff will introduce nonlinear distortions that reduce signal quality, so many approaches have been proposed to combat this problem.
For example, a common strategy in \ac{OFDM} systems is to reduce the \ac{PAPR}\cite{fink2018extrapolated}, because lowering the \ac{PAPR} reduces the nonlinear distortion of the \ac{PA}.
To this end, 5G systems (optionally) employ \ac{DFT-s-OFDM}.
However, nonlinear distortions cannot be completely avoided with standard methods for \ac{PAPR} reduction, so additional techniques to mitigate nonlinear distortions of \acp{PA} are required.

In particular, \ac{DPD} has been utilized as a method for mitigating non-linear distortions\cite{katz2016evolution, morgan2006generalized}.
This approach typically requires a feedback loop for estimating \ac{PA} characteristics and involves sophisticated signal processing at the transmitter.
In the uplink, this increases complexity, cost, and power usage in the \ac{UE}.
Therefore, to avoid complex approaches at low cost transmitters, we can use \ac{DPoD}, which compensates for non-linear distortions of uplink transmissions at the receiver, leveraging its large computational resources\cite{tellado2003maximum}.
Consequently, in recent years a great deal of effort has been devoted to \ac{DPoD} approaches, such as those in \cite{tehrani2010successive}, which utilizes a successive cancellation scheme to minimize interference from adjacent users in a multiuser scenario.
An alternative approach is discussed in \cite{alina2015digital}, which utilizes nonlinear components instead of compensating for them to increase performance, while assuming the \ac{PA} characteristics are known at the receiver.
These studies predominantly employ traditional \ac{OFDM} rather than \ac{DFT-s-OFDM}.
Noteworthy exceptions include \cite{farhadi2023deep}, where a neural network is applied in place of a conventional demapper, and \cite{babaroglu2024digital}, where the authors utilize memory polynomials and enhance the pilot sequence to mitigate nonlinear distortions at the receiver.
Recently, \ac{DPoD} has been considered for future integration into upcoming 5G standard releases\cite{3gpp_postdistortion}.

Against this background, this study investigates the practicality and challenges associated with the implementation of \ac{DPoD} in the time-domain, frequency-domain, and DFT-s-domain in a 5G \ac{DFT-s-OFDM} system.
We subsequently illustrate the transformation of a complex-valued problem formulation for counteracting nonlinearity into a real-valued problem.
This technique was proposed in \cite{slavakis2009adaptive}, but the proof is only given for linear functions.
This study provides a novel proof for nonlinear functions, which highlights the structure of the resulting functions and substantiates the achieved performance enhancements.
Furthermore, we show the application of the established Volterra series for nonlinearity compensation, as well as an alternative formulation in a \ac{RKHS}, which offers superior asymptotic computational efficiency.
We show that the classical solution to the Tikhonov-regularized least squares minimization problem that arises in the \ac{RKHS} problem formulation can be interpreted as an orthogonal projection onto a closed subspace in a Cartesian product of Hilbert spaces.
We also conduct comprehensive simulations to substantiate the performance improvements achievable through the suggested algorithms.

The remainder of this paper is organized as follows.
In \cref{sec:system_model} we describe the system model.
In \Cref{sec:receiver_processing} we give a detailed analysis of the processing that is required on the receiver side to compensate for the \ac{PA} nonlinearity.
In \cref{sec:algorithm} we present the algorithms we use to remove the nonlinear distortions, and in \cref{sec:simulations} we present the simulation setup and results.

\subsection{Notation}

In the following, scalars, column vectors, and matrices are denoted by italic lowercase letters $x$, bold lowercase letters $\Vx$, and bold uppercase letters $\VX$, respectively.
Matrix transposes, hermitian transposes, and inverses are denoted by $\VX\tra$, $\VX\herm$ and $\VX\inv$, respectively.
The sets of non-negative integers, integers, real numbers and complex numbers are denoted by $\IN$, $\IZ$, $\IR$ and $\IC$, respectively, while the real and imaginary parts of a complex number $c \in \mathbb{C}$ are denoted by $\Re(c)$ and $\Im(c)$, respectively.
The identity matrix, all-zero matrix and unitary \ac{DFT} matrix (we use the convention that the zero frequency component of a vector in the frequency-domain is in the middle) are denoted by $\VI_N \in \IR^{N \times N}$, $\mathbf{0}_{N \times M} \in \IR^{N \times M}$, and $\VF_N \in \IC^{N \times N}$, respectively.
The empty set is denoted by $\emptyset$.
Element-wise multiplication, element-wise division, the Kronecker product and cyclic convolution are denoted by $\odot$, $\oslash$, $\otimes$ and $*$, respectively.
We denote by $[\Vx]_i$ the $i$th element of the vector $\Vx$, by $[\VX]_{ij}$ the element in the $i$th row and $j$th column of the matrix $\VX$, and by $\sbr{\VX}_{:,i}$ the $i$th column of the matrix $\VX$.
The trace of $\VX$ is denoted by $\tr(\VX)$.

\section{System Model} \label{sec:system_model}

We consider a transmitter compliant with 5G standards, employing \ac{DFT-s-OFDM} (also referred to as \ac{SC-FDMA} or transform precoding) with a \ac{FFT} size of $N$, where $M<N$ subcarriers are dedicated to data transmission, and the \ac{RF} chain incorporates a non-linear \ac{PA}.
The transmitted symbols for a single \ac{OFDM} symbol, after \ac{LDPC} encoding, rate matching, scrambling, and \ac{QAM} modulation, are represented as $\Vs_\Ud \in \IC^M$ (which is called the DFT-s-domain in the following).
We represent the frequency domain signal as $\Vs_\Uf = \VF_M \Vs_\Ud \in \IC^M$. Additionally, we represent its time domain counterpart as $\Vs_\Ut = \VF\inv_N \VS \Vs_\Uf \in \IC^N$.
Here, $\VS = \sbr{\mathbf{0}_{M \times g_\Ul} \VI_M \mathbf{0}_{M \times g_\Uu}}\tra \in \IR^{N \times M}$ serves as the subcarrier mapping matrix. The parameters $g_\Ul \in \IN$ and $g_\Uu \in \IN$ specify the number of lower and upper guard carriers, respectively.
We use the 3GPP \ac{GMP} \ac{PA} model \cite{pamodel}, where we clamp the magnitude of each input sample to 1 before applying the model, because the model generates implausible results for $\Norm{\Vx}_\infty > 1$.
For a given input vector $\Vx$ the output of the \ac{GMP} model is defined element-wise as
\begin{equation}
\begin{aligned}\relax
    [q(\Vx)]_n &:= \sum_{k \in K_a} \sum_{l \in L_a} a_{kl} [\Vx_\Uc]_{n-l} |[\Vx_\Uc]_{n-l}|^{2k} \\ &+ \sum_{k \in K_b} \sum_{l \in L_b} \sum_{m \in M_b} b_{klm} [\Vx_\Uc]_{n-l} |[\Vx_\Uc]_{n-l-m}|^{2k}, \label{eq:pa_model}
\end{aligned}
\end{equation}
where $\Vx \in \IC^N$ represents the complex baseband signal, $[\Vx_\Uc]_n := \frac{[\Vx]_n}{\max(|[\Vx]_n|, 1)}$ is the clamped input, $K_a \subset \IN$ is the set of polynomial orders of the diagonal terms, $K_b \subset \IN$ is the set of polynomial orders of the cross terms, $L_a \subset \IZ$ is the set of time shifts that are applied to signal and signal envelope of the diagonal terms, $L_b \subset \IZ$ is the set of time shifts that are applied to signal and signal envelope of the cross terms, $M_b \subset \IZ$ is the set of time shifts applied only to the signal envelope of the cross terms and $a_{kl} \in \IC$ and $b_{kml} \in \IC$ are the coefficients of diagonal and cross terms respectively.
Simplified variations of this model include the \ac{MP} model where $K_b = \emptyset$, i.e., there are no cross-terms, and the memoryless polynomial model where $K_b = \emptyset$ and $L_a = \cbr{0}$.

Under the assumption that the channel impulse response has a maximum delay shorter than the \ac{CP} length, each \ac{OFDM} symbol can be processed individually.
This allows us to represent the received signal in the complex baseband as $\Vx_\Ut = \Vh_\Ut * q(\Vs_\Ut) + \Vn \in \IC^N$, where $\Vh_\Ut \in \IC^\ell$ is the channel impulse response with length $\ell \in \IN$ and $\Vn \in \IC^N$ is the \ac{AWGN} with variance $\sigma_\Un^2$.


\section{Proposed Receiver Structure} \label{sec:receiver_processing}

In the following we describe the signal processing operations on the receiver side and investigate four methods of compensating for the \ac{PA} nonlinearity.
To simplify the exposition of the main ideas, we assume a noise-free scenario and a channel frequency response without zero entries, in which case channel equalization can be achieved by channel inversion (in \cref{sec:simulations} we will show simulations without these assumptions).
However, our approach can be trivially extended to the noisy case by using \ac{LMMSE} channel equalization instead.
In more detail, the four methods we consider can be summarized as follows:

\begin{figure*}[t]
    \centering
    \includegraphics[width=0.9\linewidth]{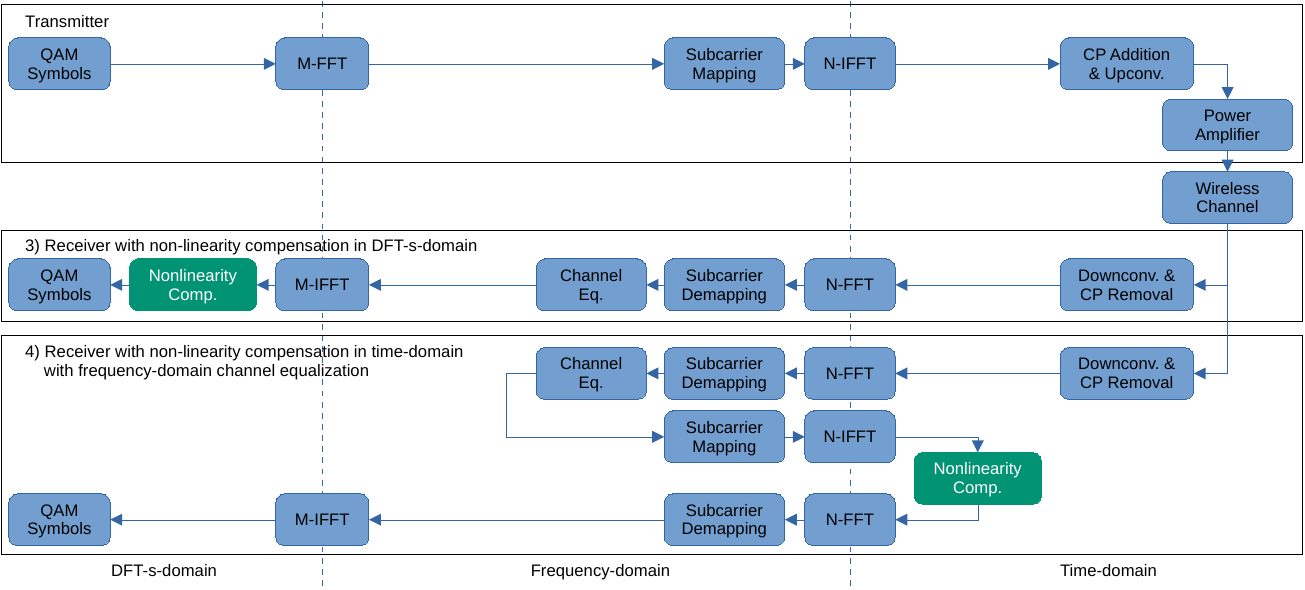}
    \caption{Architecture of transmitter, and several proposed versions of receiver. Operations that are not relevant to this work have been omitted.}
    \label{fig:algo_graph}
\end{figure*}

\begin{enumerate}
\item \emph{Time-domain nonlinearity compensation:}
The straight-forward way to estimate the transmit signal $\Vs_\Ut$ is to undo the effect of $\Vh_\Ut * q(\cdot)$.
However, because channel equalization and nonlinearity compensation is performed jointly and the channel impulse response changes fast, this necessitates rapid training with minimal training data.

\item \emph{Frequency-domain nonlinearity compensation:}
The problem of having to learn the fast-changing channel while compensating for the nonlinearity can be avoided by using traditional pilot-based channel estimation in the frequency-domain to invert the channel and then undoing the non-linearity in a second step.
In this case the symbols after subcarrier demapping can be represented as $\Vx_\Uf = \VS\tra \rbr{\Vh_\Uf \odot \VF_N q(\Vx_\Ut) \oslash \hat{\Vh_\Uf}} = \VS\tra \VF_N q(\Vx_\Ut)$, with $\hat{\Vh_\Uf} = \VF_N \hat{\Vh_\Ut}$ representing the estimated channel frequency response, where we assume $\hat{\Vh_\Uf} = \Vh_\Uf$ at this stage.
Compensating for the non-linearity in the frequency-domain presents a significant computational challenge, as the nonlinearity disperses across all subcarriers, implying the need for processing all subcarriers jointly.

\item \emph{DFT-s-domain nonlinearity compensation:}
Instead, the nonlinearity compensation can be executed in the DFT-s-domain, where the symbol is represented by $\Vx_\Ud = \VF_M\inv \VS\tra \VF_N q(\Vx_\Ut)$. This method benefits from the concentration of nonlinear distortion from a single time-domain sample into a limited number of symbols, due to $\VF_M\inv \VS\tra \VF_N$ having most of the energy around the subcarriers with row index $i$ and column index $j$, where $\frac{i}{j} \approx \frac{M}{N}$.
However, the rows of $\VF_M\inv \VS\tra \VF_N$ are generally not shifted versions of each other, which implies that each symbol is influenced by the nonlinearity in a different way.
This complicates the learning process.

\item \emph{Time-domain nonlinearity compensation with frequency-domain channel equalization:}
We propose to undo the nonlinearity in time-domain after frequency-domain channel equalization by calculating the channel-equalized time-domain symbols as $\Vx_\Uc = \VF_N\inv \VS \VS\tra \VF_N q(\Vx_\Ut)$.
Because $\VF_N\inv \VS \VS\tra \VF_N$ is a circulant matrix, this can be written in terms of a cyclic convolution as $\Vx_\Uc = \Vb * q(\Vx_\Ut)$ where $\Vb = \sbr{\VF_N\inv \VS \VS\tra \VF_N}_{:,1}$.
The vector $\Vb$ represents the impulse response of an ideal lowpass filter, which is applied to the output of the nonlinearity because the subcarriers in the guard band are zeroed out by the subcarrier demapping/mapping operation due to the lack of pilots in the guard band.
The advantage of this method over DFT-s-domain nonlinearity compensation is that each sample can be derived from a cyclic convolution with the same vector, which can be exploited by applying the same algorithm to all samples individually.
Additionally, the nonlinearity can be undone without taking the channel into account, because $\Vb$ does not depend on $\Vh_\Ut$ (which is not the case for direct time-domain nonlinearity compensation).
\end{enumerate}

An overview of the different methods is shown in \cref{fig:algo_graph}.
In the remainder of this study we will focus exclusively on options 3) and 4) because we consider options 1) and 2) to have prohibitive computational requirements for real-time application.

\section{Algorithm} \label{sec:algorithm}

In this section, we first show how to formulate the complex-valued problem of nonlinearity compensation as a real-valued problem.
Then we present two distinct methods to solve the real-valued problem: the Volterra series and a kernel method employing polynomial kernels.
Subsequently, we analyze their respective complexities. 

\subsection{Problem Formulation in Real Hilbert Spaces} \label{sec:complex_to_real}

By exploiting the fact that the nonlinearities in \eqref{eq:pa_model} satisfy $q(\Uj\Vx) = \Uj q(\Vx)$, we rephrase the problem of finding a complex-valued function $g: \IC^N \to \IC$ to the problem of finding a real-valued function $f: \IR^{2N} \to \IR$ able to recover the real and imaginary parts of $g$ by simply rearranging the elements of the input vector to $f$.
To this end, we assume that $g$ can be expressed as $(\forall \Vx \in \IC^N) g(\Vx) = f(\Vx_1) + \Uj f(\Vx_2)$, with $\Vx_1 = \sbr{\Re(\Vx)\tra, \Im(\Vx)\tra}\tra \in \IR^{2N}$ and $\Vx_2 = \sbr{\Im(\Vx)\tra, -\Re(\Vx)\tra}\tra \in \IR^{2N}$.
This assumption reduces the search space to complex functions $g$ satisfying $g(\Uj\Vx) = \Uj g(\Vx)$ (such as in \eqref{eq:pa_model}), which enables us to exploit model knowledge in the proposed algorithm described later simply by applying the algorithm in real domain instead of complex domain.
To prove this equivalence we first present the following fact.
\begin{fact}[{\hspace{1sp}\cite[Proposition 5.1]{fink2022fixed}}]\label{fact:complex_real_iso}
The real Hilbert space $\rbr{\IC^N, \abr{\cdot, \cdot}_\IC, \Norm{\cdot}_\IC}$ equipped with the inner product $\abr{\Vx, \Vy}_\IC := \Re\rbr{\Vy\herm\Vx}$ is isometrically isomorphic to the real Hilbert space $\rbr{\IR^{2N}, \abr{\cdot, \cdot}, \Norm{\cdot}}$ with the standard Euclidean inner product.
This isomorphism is achieved through the bijective linear mapping $\xi: \IC^N \to \IR^{2N}$, which is defined by $\xi(\Vx) := \sbr{\Re(\Vx)\tra, \Im(\Vx)\tra}\tra$.
\end{fact}

This enables us to obtain the following result.
\begin{proposition}
    A function $g: \IC^N \to \IC$ satisfies $g(\Uj \Vx) = \Uj g(\Vx)$ if and only if there exists an odd function $f: \IR^{2N} \to \IR$, such that $g(\Vx) = f(\Vx_1) + \Uj f(\Vx_2)$, where $\Vx_1 = \sbr{\Re(\Vx)\tra, \Im(\Vx)\tra}\tra \in \IR^{2N}$ and $\Vx_2 = \sbr{\Im(\Vx)\tra, -\Re(\Vx)\tra}\tra \in \IR^{2N}$.
\end{proposition}

\emph{Proof:} Assume that $g$ satisfies $g(\Uj \Vx) = \Uj g(\Vx)$.
We note that $g(\Uj\Vx) = \Uj g(\Vx)$ implies $g(-\Vx) = g(\Uj\Uj\Vx) = \Uj\Uj g(\Vx) = -g(\Vx)$ and $g(-\Uj \Vx) = -\Uj g(\Vx)$.
As a result, $g(\Vx) = \Re(g(\Vx)) + \Uj\Im(g(\Vx)) \overset{\mathrm{(a)}}{=} \Re(g(\Vx)) + \Uj\Re(-\Uj g(\Vx)) = \Re(g(\Vx)) + \Uj\Re(g(-\Uj\Vx))$, where (a) follows from $\Im(\Vx) = \Re(-\Uj\Vx)$.
In light of \cref{fact:complex_real_iso}, we define a function $f: \IR^{2N} \to \IR : \Vx_1 \mapsto \Re(g(\xi^{-1}(\Vx_1)))$, which allows us to express $g(\Vx) = f(\xi(\Vx)) + \Uj f(\xi(-\Uj \Vx)) = f(\Vx_1) + \Uj f(\Vx_2)$.
In addition, $f$ is odd because $g$ is odd.


Conversely, assume that there exists an odd function $f$ such that $g(\Vx) = f(\Vx_1) + \Uj f(\Vx_2)$.
This allows us to prove with simple algebraic manipulations that $g(\Vx)$ satisfies $g(\Uj \Vx) = \Uj g(\Vx)$.
Indeed, $g(\Uj\Vx) = f\rbr{\sbr{\Re(\Uj\Vx)\tra, \Im(\Uj\Vx)\tra}\tra} + \Uj f\rbr{\sbr{\Im(\Uj\Vx)\tra, -\Re(\Uj\Vx)\tra}\tra} \overset{\mathrm{(a)}}{=} f\rbr{\sbr{-\Im(\Vx)\tra, \Re(\Vx)\tra}\tra} + \Uj f\rbr{\sbr{\Re(\Vx)\tra, \Im(\Vx)\tra}\tra} = f(-\Vx_2) + \Uj f(\Vx_1) \overset{\mathrm{(b)}}{=} \Uj (f(\Vx_1) -\Uj f(-\Vx_2)) = \Uj (f(\Vx_1) + \Uj f(\Vx_2)) = \Uj g(\Vx)$, where (a) follows from $\Re(\Uj\Vx) = -\Im(\Vx)$ and $\Im(\Uj\Vx) = \Re(\Vx)$ and (b) follows from multiplying by $-\Uj\Uj=1$. $\hfill\blacksquare$

The initial suggestion for this proposition has been given in \cite{slavakis2009adaptive}, albeit specifically for linear functions.

To avoid confusion we will denote complex vectors with underscore and their real counterpart without underscore in the remainder of this section.

\subsection{Volterra Series} \label{sec:volterra_series}

Next we present the first method for mitigating nonlinear distortions.
In this method, we solve $\underline{\Vy} = \Vb * q(\underline{\Vx})$ for $\underline{\Vx}$, where $\underline{\Vx}, \underline{\Vy} \in \IC^N$.
We note that a memoryless nonlinearity filtered with a low-pass filter is a special case of a nonlinearity with memory.
The nonlinear distortions will be removed from each element of $\underline{\Vx}$ independently, so we need to define the input with memory that is used to predict each output sample.
Owing to the cyclic nature of the convolution, the nonlinearity's memory is cyclic as well.
We exploit this fact by using $\underline{\Vy}_n \in \IC^L$ as input to predict $[\underline{\Vx}]_n$, where $\underline{\Vy}_n$ is defined element-wise as $[\underline{\Vy}_n]_i := [\underline{\Vy}]_{(n-[\Vell]_i) \bmod N}$, with $\Vell \in \IZ^L$ being the vector of memory time shifts and $L \in \IN$ being the memory depth.
Following the argument from \cref{sec:complex_to_real} we work with the real-valued vectors $\Vx = \xi(\underline{\Vx}) \in \IR^{2N}$ and $\Vy_n = \xi(\underline{\Vy}_n) \in \IR^{2L}$ in the the remainder of this section.

In the following, we assume that there are $K$ training samples with known transmit samples at hand.
These could be gathered offline in a testing environment, or alternatively, data can be sent with a sufficiently low \ac{MCS} level that allows for recovering the transmitted data even in the presence of nonlinear distortion.
Given the known transmitted symbols $\Vs_\Ud$ and received channel-equalized frequency-domain signal $\Vx_\Uf$ we can derive the training data as $\underline{\Vx} = \Vs_\Ud$ and $\underline{\Vy} = \VF_M\inv \Vx_\Uf$ for nonlinearity compensation in the DFT-s-domain and as $\underline{\Vx} = \VF_N\inv \VS \VF_M \Vs_\Ud$ and $\underline{\Vy} = \VF_N\inv \VS \Vx_\Uf$ for nonlinearity compensation in the time-domain.

Nonlinearities can be represented by Volterra series where only terms with odd order are considered, because terms with even orders only generate out-of-band distortion (under the assumption that the bandwidth of the signal is much smaller than the carrier frequency)\cite{morgan2006generalized}.
For this reason the (real) Volterra series of interest have the structure
\begin{equation}
    [\hat{\Vx}]_n = \sum_{k\in \CD} \sum_{m_1=1}^{2L} \cdots \sum_{m_k=1}^{2L} h_k(m_1, \ldots, m_k) \prod_{l=1}^k [\Vy_n]_{m_l}, \label{eq:volterra_series}
\end{equation}
where $\CD = \cbr{1, 3, \ldots, d}$ is the set of odd integer polynomial powers up to degree $d \in \IN$ and $h_k(m_1, \ldots, m_k) \in \IR$ is the Volterra kernel, which needs to be estimated.
The Volterra series in \eqref{eq:volterra_series} can be written more compactly and without redundant terms as $[\hat{\Vx}]_n = \Vh\tra \Va_n$, where $\Vh \in \IR^P$ with $P = \sum_{k \in \CD} \binom{k + 2L - 1}{2L}$ is the vectorized Volterra kernel and $\Va_n \in \IR^P$
is the the vector of all monomials of the entries of $\Vy_n$ with degree in $\CD$.
We note that the \ac{GMP}, \ac{MP} and memoryless polynomial models are all special cases of the Volterra series.

The Volterra kernel corresponds to a solution to the optimization problem
\begin{equation*}
    \minimize_{\Vh \in \IR^P} \sum_{i=1}^K \rbr{\Vh\tra \Va_i - [\Vx]_i}^2,
\end{equation*}
which has the well-known least squares solution
\begin{equation*}
    \Vh = \rbr{\VA\tra \VA}\inv \VA\tra \Vx,
\end{equation*}
where $\VA = \sbr{\Va_1, \ldots,  \Va_K} \in \IR^{K \times P}$ (see also \cite{babaroglu2023cellular}).

\subsection{Kernel Method}

As shown in \cite{franz2006unifying}, the Volterra series can be represented implicitly as an element of a RKHS $(\CH_\UK, \abr{\cdot, \cdot}_{\CH_\UK}, \Norm{\cdot}_{\CH_\UK})$ induced by a kernel function $\kappa: (\Vu, \Vv) \mapsto \sum_{k \in \CD} (\Vu\tra \Vv)^k$\cite[Chapter 1]{berlinet2011reproducing}. 
This property allows us to devise an algorithm aiming to compensate for the nonlinearity via an optimization problem directly within the \ac{RKHS}.
More specifically, the objective is to obtain a function $f: \IR^{2L} \to \IR$ from the \ac{RKHS} that strikes a balance between data fidelity and model simplicity.
This function is used to estimate the samples without nonlinearity as $[\hat{\Vx}]_n = f(\Vy_n)$.
More formally the problem can be stated as
\begin{equation}
    \minimize_{f\in \CH_K} \sum_{i=1}^K \rbr{f(\Vy_i) - [\Vx]_i}^2 + \lambda \Norm{f}^2_{\CH_K}, \label{eq:rkhs_closed_form_problem}
\end{equation}
where $\lambda > 0 \in \IR$ is a regularization parameter that trades data fidelity against model simplicity.
The solution to this problem is shown below.

\begin{proposition} \label{prop:rkhs_closed_form}
    The unique minimizer of \eqref{eq:rkhs_closed_form_problem} is given by
    \begin{equation*}
        f^\star = \sum_{i=1}^K [\boldsymbol{\beta}]_i \kappa(\Vy_i, \cdot)
    \end{equation*}
    with
    \begin{equation*}
        \boldsymbol{\beta} = \rbr{\VK + \lambda \VI}\inv \Vy,
    \end{equation*}
    where $\VK \in \IC^{K \times K}$ is the Gram matrix defined element-wise by $[\VK]_{ij} = \kappa(\Vy_i, \Vy_j)$.
\end{proposition}
The result in \cref{prop:rkhs_closed_form} is known from \cite{gockenbach2016linear} in slightly different form. A detailed proof based on \cite{renato_lecture} is given in the appendix for completeness.

\subsection{Complexity Analysis}

For predicting a single sample, the Volterra series approach requires $\CO\rbr{\binom{d+L}{L}}$ multiplications, whereas the \ac{RKHS} approach requires $\CO(K(L+d))$ multiplications.
Required storage is $\CO\rbr{\binom{d+L}{L}}$ elements for Volterra series and $\CO(KL)$ elements for \ac{RKHS}.
This illustrates that \ac{RKHS} offers lower complexity when dealing with high memory depth or polynomial degree.
Conversely, the Volterra series implementation's complexity does not depend on the number of training samples, making it preferable for scenarios with a large number of training samples.
For comparison, the \ac{MP} approach from \cite{babaroglu2024digital} requires $\CO\rbr{d^2 L}$ multiplications per sample.

The training complexity of the Volterra series is given by
\begin{equation*}
    \CO\rbr{\max\rbr{K, \binom{d+L}{L}}^2 \min\rbr{K, \binom{d+L}{L}}},
\end{equation*}
whereas the RKHS method has a complexity of $\CO\rbr{K^3 + K^2(L+d)}$, which can be prohibitive for time-constrained environments.
However, iterative techniques tend to reduce computational demands significantly.
The development of appropriate algorithms will be addressed in future research.
Additional efficiency improvements can be achieved by utilizing techniques like coefficient pruning\cite{faucris.110099484}.

\section{Simulations} \label{sec:simulations}

To assess the efficacy of the proposed algorithms, we perform 5G \ac{PUSCH} simulations utilizing an adapted version of Sionna\cite{sionna}, which includes several enhancements and corrections.\footnote{The source code can be found at \url{https://github.com/danielschaeufele/sionna}}
Details on the simulation setup are provided in \cref{tab:simulation_parameters}.
The training dataset was generated by employing an \ac{AWGN} channel with a \ac{SNR} of \SI{50}{dB}.
In practice this could be achieved in a controlled laboratory environment where the measurement devices are attached to the \ac{PA} with a cable.
The training dataset is composed of 4 \ac{OFDM} symbols, resulting in $4N = 16384$ training samples for time-domain compensation and $4M = 12960$ samples for DFT-s-domain compensation.
When using the \ac{GMP} model\cite{pamodel}, we upsample the time-domain signal by a factor of 3 by zero-padding the input to the \ac{DFT}, aiming to closely match the nominal sample rate of \SI{400}{MHz}.
All algorithms assume a polynomial degree $d=5$ and we use the memory indices $\Vell = \sbr{0}$, $\Vell = \sbr{-2, \ldots, 2}\tra$ and $\Vell = \sbr{-5, \ldots, 0}\tra$ for the algorithms labelled \emph{no memory}, \emph{symmetric memory} and \emph{asymmetric memory} respectively.
These values were chosen to provide a good compromise between computational complexity and \ac{BER} performance.
When performing \ac{DPoD} in time-domain with symmetric memory we opt for the kernel method due to its lower complexity.
However, we confirmed numerically that the Volterra technique has identical performance.
The regularization parameter for the kernel method was chosen to be $\lambda = 0.005 \frac{\tr(\VK)}{K}$ by simple line search.
To achieve an accurate comparison with \cite{babaroglu2024digital}, we employ the same training dataset to train their algorithm, omitting their pilot boosting technique.

\begin{table}[tb]
    \centering
    \begin{tabular}{cc}
    \toprule
         \textbf{Parameter} & \textbf{Value} \\
    \midrule
         FFT size $N$ & 4096 \\
         DFT size $M$ & 3240 \\
         Subcarrier spacing & \SI{30}{kHz} \\
         Bandwidth & \SI{100}{MHz} \\
         Sample rate & \SI{122.88}{MHz} \\
         Number of Physical Ressource Blocks & 270 \\
         MCS level & 26 \\
         Carrier frequency & \SI{28}{GHz} \\
         Channel model & TDLD30-75\cite{3gpp_channel_model} \\
         Delay spread & \SI{30}{ns} \\
         Maximum Doppler frequency & \SI{75}{Hz} \\
         PA model & $\sim$\SI{28}{GHz}, GaN \cite{pamodel} \\
         Power backoff & \SI{6}{dB} \\
    \bottomrule
    \end{tabular}
    \vspace{2mm}
    \caption{Simulation parameters}
    \label{tab:simulation_parameters}
\end{table}

\begin{figure}[tb]
    \centering
    \includegraphics[width=1\linewidth]{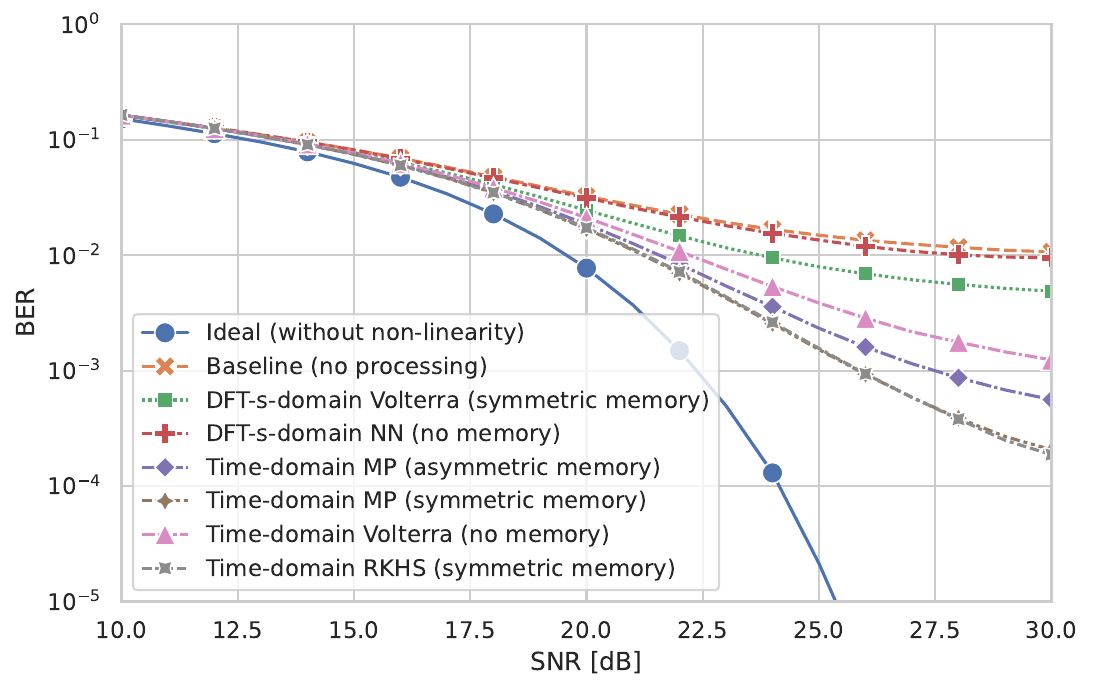}
    \caption{Simulation Results (uncoded \ac{BER}) for \ac{AWGN} channel and memoryless polynomial \ac{PA}}
    \label{fig:results_ber_simple}
\end{figure}

The results of simulations with the memoryless polynomial model and \ac{AWGN} channel model are shown in \cref{fig:results_ber_simple}.
As expected, the time-domain nonlinearity compensation surpasses its DFT-s-domain counterpart.
Furthermore, an algorithm that incorporates memory demonstrates significantly improved performance, even when using a memoryless polynomial model for the \ac{PA}, confirming our prior analysis in \cref{sec:receiver_processing}.
The algorithm developed in \cite{farhadi2023deep} (termed \emph{DFT-s-domain NN (no memory)}) does not perform satisfactorily, as the approach does not take into account the aforementioned insights.
The approach suggested in \cite{babaroglu2024digital} utilizes memory extending only backward in time, which cannot approximate the convolution with an ideal lowpass filter.
Consequently, our proposed algorithm outperforms theirs (termed \emph{Time-domain MP (asymmetric memory)}).
When changing their algorithm to use symmetric memory (termed \emph{Time-domain MP (symmetric memory)}), its performance matches that of our proposed method.

\begin{figure}[tb]
    \centering
    \includegraphics[width=1\linewidth]{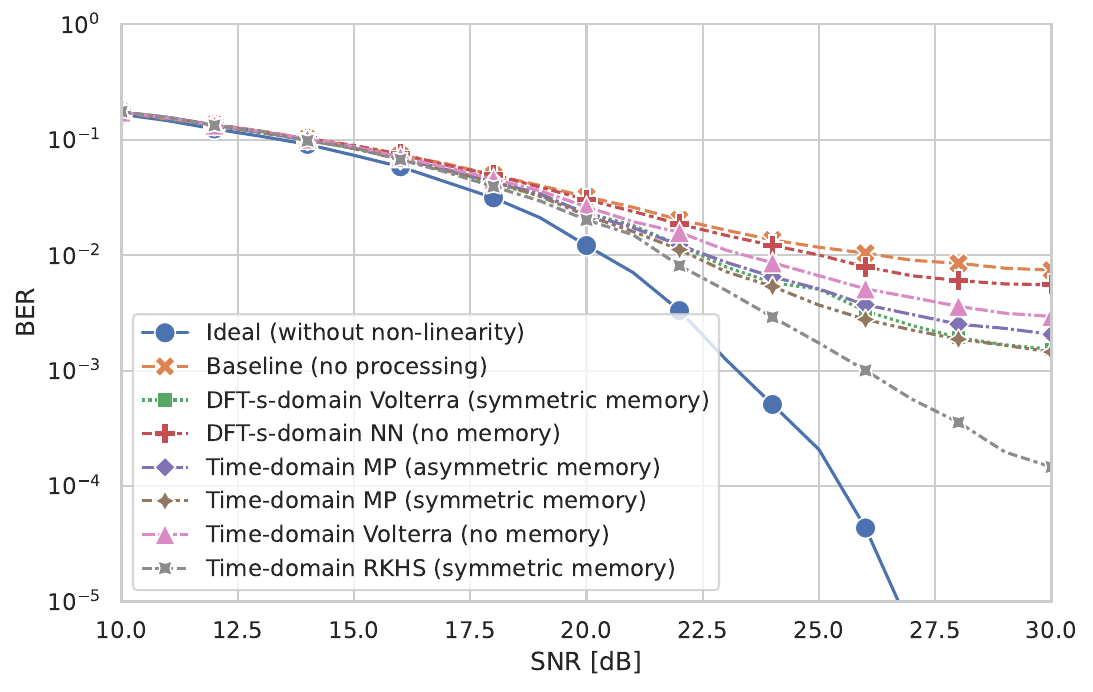}
    \caption{Simulation Results (uncoded \ac{BER}) for TDL-D channel and \ac{GMP} \ac{PA}}
    \label{fig:results_ber_complex}
\end{figure}

\begin{figure}[tb]
    \centering
    \includegraphics[width=1\linewidth]{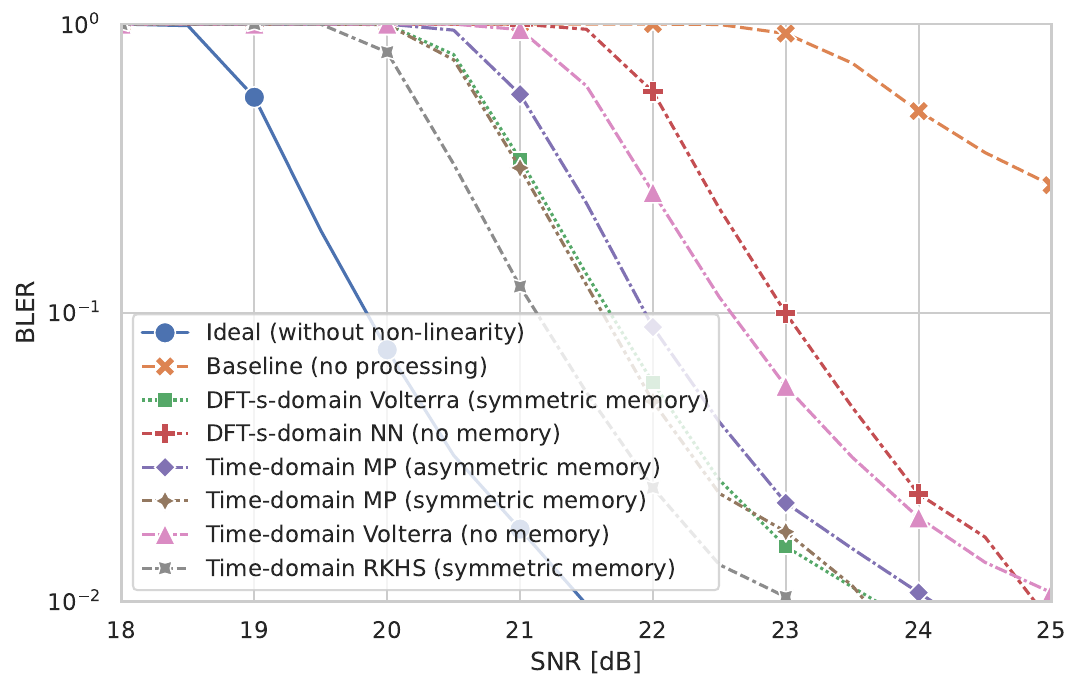}
    \caption{Simulation Results (\ac{BLER}) for TDL-D channel and \ac{GMP} \ac{PA}}
    \label{fig:results_bler_complex}
\end{figure}

Subsequently, we demonstrate that our algorithm exhibits greater robustness across various \ac{PA} models.
As shown in \cref{fig:results_ber_complex}, simulations utilizing the TDL-D channel model and the \ac{GMP} \ac{PA} model reveal that our method outperforms the algorithm outlined in \cite{babaroglu2024digital} even when using symmetric memory.
This stems from the presence of cross-terms in the \ac{GMP} model, which cannot be adequately counteracted by the \ac{MP} model, hence establishing the superior resilience of the Volterra model in handling complex \ac{PA} characteristics.
\Cref{fig:results_bler_complex} shows the \ac{BLER} with \ac{LDPC} coding as specified in the 5G standard.
Consequently, these results indicate that the proposed algorithm achieves an \ac{SNR} gain of \SI{1}{dB} over the method from \cite{babaroglu2024digital} and a \SI{2}{dB} gain over \cite{farhadi2023deep} at the \SI{10}{\percent} \ac{BLER} threshold.
When compared to a baseline without nonlinearities, there is only a \SI{1.5}{dB} \ac{SNR} reduction, showcasing that integrating a cost-effective, energy-saving \ac{PA} with \ac{DPoD} is a practical alternative to using an over-provisioned \ac{PA} with high backoff and low nonlinear distortion.


\section{Conclusion}

In this study, we explored multiple approaches for mitigating the nonlinearity of \ac{PA} on the receiver end.
Initially, we analyzed the pros and cons of compensating nonlinearity in the time-domain, frequency-domain, and DFT-s-domain.
During this analysis, we observed that taking memory effects into account is beneficial, even when using a memoryless \ac{PA} model.
Subsequently, we introduced a method for imposing a particular structure on complex functions by expressing them via real functions and provided a novel proof for nonlinear functions, showcasing the prerequisites and advantages of this approach when used for estimating nonlinear functions.
We then discussed the traditional Volterra series and proposed an equivalent kernel method that can significantly reduce algorithmic complexity.
Finally, we derived a closed-form solution to the regularized least squares problem in a \ac{RKHS}.
Simulations show that our proposed algorithm can significantly improve performance compared to state-of-the-art algorithms and that \ac{DPoD} is a suitable method to deal with the impairments that are caused by using cheaper and more energy-efficient \acp{PA}, while moving the computational cost to the base station side, where compute resources are more ubiquitous.


In forthcoming studies, we intend to explore iterative techniques such as \ac{SGD} within the Volterra series framework and \ac{APSM}\cite{yamada2005adaptive} in the context of kernel methods.
Additionally, we plan to conduct empirical measurements utilizing an actual hardware \ac{PA}.

\appendix

\subsection{Proof of \cref{prop:rkhs_closed_form}}

In the following, we derive the solution to problem \eqref{eq:rkhs_closed_form_problem} in closed form as the projection of a specific point onto a closed convex set in a Hilbert space.
To see the estimation problem as a projection, we consider the vector space defined as a Cartesian product of Hilbert spaces $\CH_\UC = \CH_\UK \times \CH$, where $\CH_\UK$ is the RKHS (to avoid confusion, we denote the inner product in $\CH_\UK$ by $\abr{\cdot, \cdot}_{\CH_\UK}$) and $\CH = \IR^K$ is the finite dimensional Hilbert space with Euclidean inner product $\abr{\Vu, \Vv} = \Vu\tra \Vv$ and norm $\Norm{\cdot} = \sqrt{\abr{\cdot, \cdot}}$, i.e. vectors in $\CH_\UC$ take the form $(f, \Vu)$, with $f \in \CH_\UK, \Vu \in \IR^K$.
To make $\CH_\UC$ a Hilbert space, we define the inner product $\abr{(f, \Vu), (g, \Vv)}_{\CH_\UC} := \lambda \abr{f, g}_{\CH_\UK} + \abr{\Vu, \Vv}$ with $\lambda > 0$ and the induced norm $\Norm{\cdot}_{\CH_\UC} = \sqrt{\abr{\cdot, \cdot}_{\CH_\UC}}$.

We first reformulate \eqref{eq:rkhs_closed_form_problem} as 
\begin{subequations} \label{eq:problem9}
\begin{alignat}{4}
&\minimize_{f \in \CH_\UK, \Vz \in \IR^K}    &\qquad& \Norm{\Vz - \Vx}^2 + \lambda \Norm{f}^2_{\CH_\UK}&\\
&\text{subject to} &      & f(\Vy_i) = [\Vz]_i,\qquad i=1,\ldots,K.&
\end{alignat}
\end{subequations}
By exploiting the reproducing property of \ac{RKHS}, we verify that \eqref{eq:problem9} is equivalent to the following optimization problem:
\begin{subequations} \label{eq:projection_problem}
\begin{alignat}{4}
&\minimize_{f \in \CH_\UK, \Vz \in \IR^K}    &\qquad& \Norm{(f, \Vz) - (\mathbf{0}, \Vx)}^2_{\CH_\UC}&\\
&\text{subject to} &      & \abr{\Vb_i, (f, \Vz)}_{\CH_\UC} = 0 \qquad i=1,\ldots,K,&
\end{alignat}
\end{subequations}
where $\Vb_i := \rbr{\kappa(\Vy_i, \cdot), -\lambda \Ve_i}$ and $\Ve_i \in \IR^K$ are the Cartesian unit vectors that have a 1 at position $i$ and are otherwise all 0.
From the definition of projections in Hilbert spaces, the solution to \eqref{eq:projection_problem} is precisely the projection from $(\mathbf{0}, \Vx)$ onto the closed subspace $M^\perp = \cbr{\Vv \in \CH_\UC | (\forall \Vw \in M) \abr{\Vv, \Vw} = 0}$, where $M = \spn\rbr{\cbr{\Vb_1, \ldots, \Vb_K}}$.

By exploiting the orthogonal decomposition theorem, we deduce $P_{M^\perp}((\mathbf{0}, \Vx)) = (\mathbf{0}, \Vx) - P_{M}((\mathbf{0}, \Vx))$, where $P_M((\mathbf{0}, \Vx))$ is derived by \cite[Chapter 3.6]{luenberger1969optimization}. 
In more detail, the projection is given by $P_M((\mathbf{0}, \Vx)) = \sum_{i=1}^K [\boldsymbol{\alpha}]_i \Vb_i$, where $\boldsymbol{\alpha} \in \IR^K$ is a solution to the normal equations
\begin{equation} \label{eq:normal_equations}
\resizebox{0.89\hsize}{!}{$
    \begin{bmatrix}
    \abr{\Vb_1, \Vb_1}_{\CH_C} & \dots & \abr{\Vb_K, \Vb_1}_{\CH_C} \\
    \vdots & \ddots & \vdots \\
    \abr{\Vb_1, \Vb_K}_{\CH_C} & \dots & \abr{\Vb_K, \Vb_K}_{\CH_C}
    \end{bmatrix}
    \begin{bmatrix}
        [\boldsymbol{\alpha}]_1 \\
        \vdots \\
        [\boldsymbol{\alpha}]_K
    \end{bmatrix}
    =
    \begin{bmatrix}
        \abr{(\mathbf{0}, \Vx), \Vb_1}_{\CH_C} \\
        \vdots \\
        \abr{(\mathbf{0}, \Vx), \Vb_K}_{\CH_C}
    \end{bmatrix}
$}
\end{equation}
We note that \eqref{eq:normal_equations} always has a solution because projections onto closed subspaces exist and are unique.
In this particular case, $\boldsymbol{\alpha}$ is unique because $\rbr{\lambda \VK + \lambda^2 \VI}$ is positive definite.

After computing the inner products, we get
\begin{equation}
    \rbr{\lambda \VK + \lambda^2 \VI} \sbr{[\boldsymbol{\alpha}]_1, \ldots, [\boldsymbol{\alpha}]_K}\tra = - \lambda\sbr{[\Vx]_1, \ldots, [\Vx]_K}\tra,
\end{equation}
where $\VK \in \IR^{K \times K}$ is the Gram matrix with entries $[\VK]_{i,j} = \kappa(\Vy_i, \Vy_j)$.

Finally, from $(f^\star, \Vz^\star) = P_{M^\perp}((\mathbf{0}, \Vx)) = (\mathbf{0}, \Vx) - P_{M}((\mathbf{0}, \Vx))$ we obtain the desired result $f^\star = \sum_{i=1}^K [\boldsymbol{\beta}]_i \kappa(\Vy_i, \cdot)$ where $\boldsymbol{\beta} := \rbr{\VK + \lambda \VI}\inv \Vx = -\boldsymbol{\alpha}$. $\hfill\blacksquare$


\bibliography{bibliography.bib}
\bibliographystyle{ieeetr}

\end{document}

%% file: acronyms.tex
\usepackage[printonlyused,nolist]{acronym}

\AtBeginDocument{
\begin{acronym}
    \acro{APSM}{adaptive projected subgradient method}
    \acro{AWGN}{additive white Gaussian noise}
    \acro{BER}{bit error ratio}
    \acro{BLER}{block error ratio}
    \acro{CAZAC}{constant amplitude zero autocorrelation}
    \acro{CP}{cyclic prefix}
    \acro{CSI}{channel state information}
    \acro{DFT}{discrete Fourier transform}
    \acro{DFT-s-OFDM}[DFT-s-OFDM]{discrete Fourier transform-spread-orthogonal frequency division multiplexing}
    \acro{DPD}{digital pre-distortion}
    \acro{DPoD}{digital post-distortion}
    \acro{FFT}{fast Fourier transform}
    \acro{GMP}{generalized memory polynomial}
    \acro{IDFT}{inverse discrete Fourier transform}
    \acro{IFFT}{inverse fast Fourier transform}
    \acro{IQ}{in-phase and quadrature}
    \acro{LDPC}{low-density parity-check}
    \acro{LLR}{log-likelihood ratio}
    \acro{LMMSE}{linear minimum mean square error}
    \acro{MCS}{modulation and coding scheme}
    \acro{MP}{memory polynomial}
    \acro{PA}{power amplifier}
    \acro{PAPR}{peak-to-average power ratio}
    \acro{PUSCH}{physical uplink shared channel}
    \acro{OFDM}{orthogonal frequency division multiplexing}
    \acro{QAM}{quadrature amplitude modulation}
    \acro{RF}{radio-frequency}
    \acro{RKHS}{reproducing kernel Hilbert space}
    \acro{SC-FDMA}{single carrier frequency division multiple access}
    \acro{SNR}{signal-to-noise ratio}
    \acro{SGD}{stochastic gradient descent}
    \acro{UE}{user equipment}
    \acro{ZC}{Zadoff-Chu}
\end{acronym}
}